\pdfoutput=1
\documentclass[a4paper]{jpconf}
\usepackage{graphicx}
\usepackage{iopams}
\begin{document}
\title{Development of high-rate capable and ultra-low mass Resistive Plate Chamber with Diamond-Like Carbon}

\author{Atsushi~Oya\textsuperscript{1}, Kei~Ieki\textsuperscript{2}, Atsuhiko~Ochi\textsuperscript{3}, Rina~Onda\textsuperscript{1}, Wataru~Ootani\textsuperscript{2} and Kensuke~Yamamoto\textsuperscript{1}}

\address{$^1$ Department of Physics, The University of Tokyo, Bunkyo-ku, Tokyo 113-0033, Japan}
\address{$^2$ International Center for Elementary Particle Physics, The University of Tokyo, Bunkyo-ku, Tokyo 113-0033, Japan}
\address{$^3$ Department of Physics, Kobe University, Kobe 657-8501 Japan}

\ead{atsushi@icepp.s.u-tokyo.ac.jp}

\begin{abstract}
A new type of resistive plate chamber (RPC) is under development using thin-film resistive electrodes based on diamond-like carbon (DLC). 
Planned to be put on the path of high-intensity low-momentum muon beam of the MEG II experiment, this detector is required to be high-rate capable and ultra-low mass.
Using a prototype detector with $2\,\mathrm{cm}\times 2\,\mathrm{cm}$ size and $0.1\%\,X_0$ material budget, performance studies were conducted for MIP detection efficiency, timing resolution and high rate capability in low-momentum muon beam.
In this paper, the measured performance is presented including the result with low-momentum muon beam at rate up to $1\,\mathrm{MHz/cm^2}$.
Based on the result, the expected performance of the full-scale detector in the MEG~II experiment is also discussed.
\end{abstract}

\section{Introduction}
A novel background identification detector is under development for the MEG II experiment, aiming for sensitivity improvement in the $\mu\to e\gamma$ decay search \cite{MEGIIdesign}. 
This detector is required to detect positrons ($<5\,\mathrm{MeV}$) from the radiative muon decays with high energy photons observed in the photon detector.
As this detector is planned to be put in a high-intensity  low-momentum muon beam with $4\,\mathrm{MHz/cm^2}$ rate at the beam center, ultra-low mass ($<0.1\,\%X_0$) and high rate capability are necessary in addition to a high efficiency ($>90\,\%$) for minimum ionizing particles (MIP) and a good timing resolution ($<1\,\mathrm{ns}$).

\begin{figure}[tbp]
\centering
\begin{minipage}{0.45\linewidth}
\centering
\includegraphics[width=0.9\linewidth]{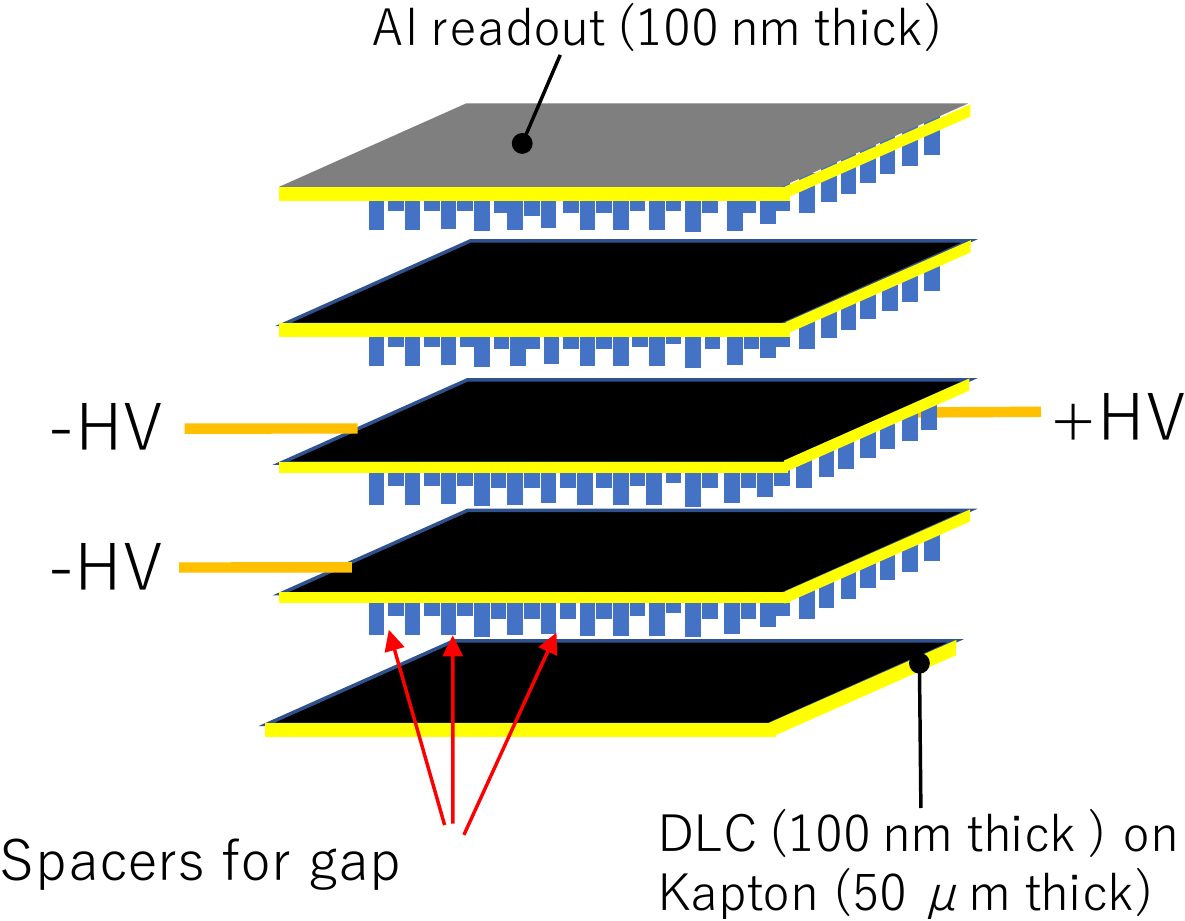}
\caption{\label{fig:DetectorConcept}Concept of the detector design.}
\end{minipage}\hspace{0.1\linewidth}%
\begin{minipage}{0.45\linewidth}
\includegraphics[width=0.8\linewidth]{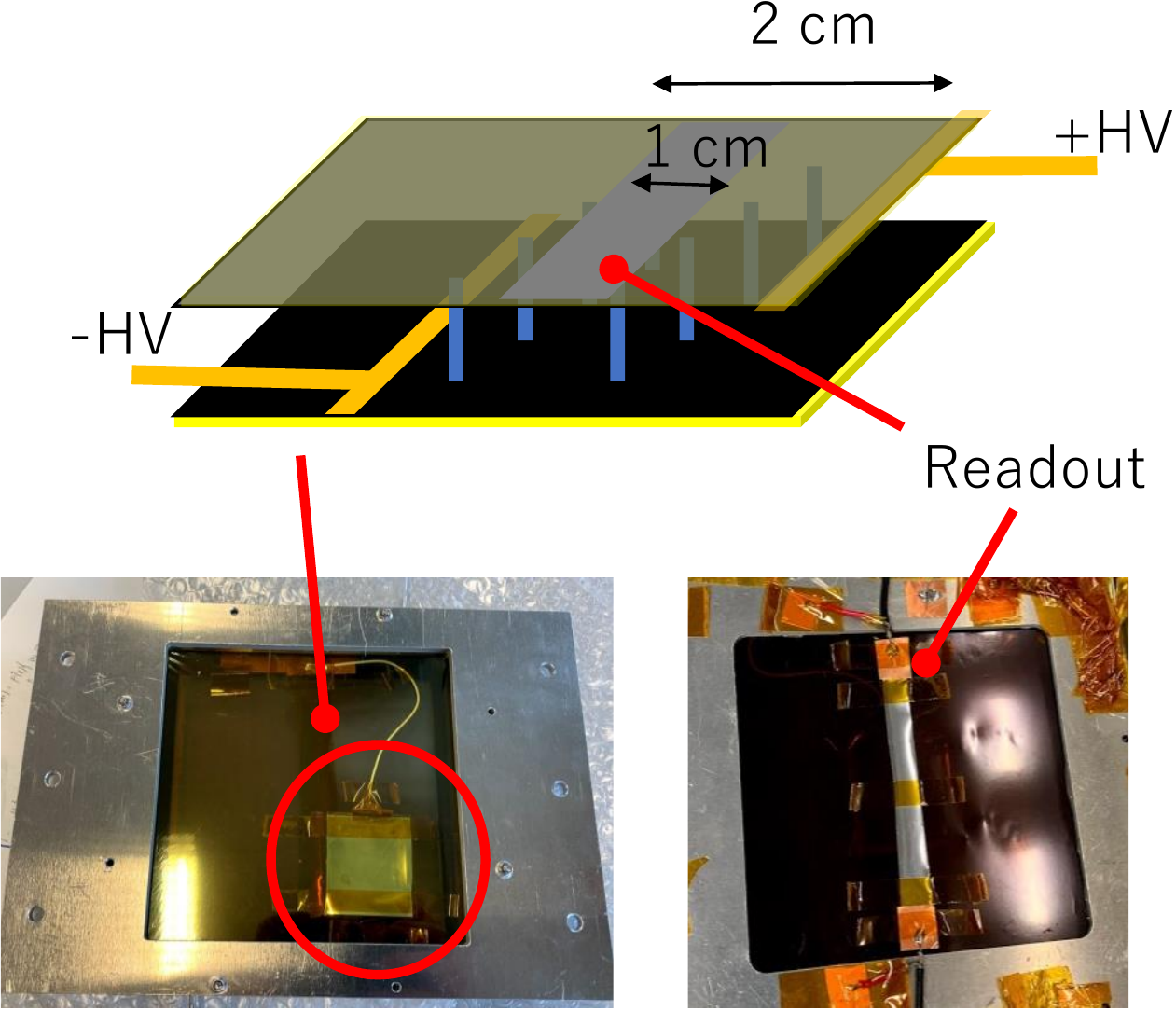}
\caption{\label{fig:PrototypePicture}Single-layer prototype.}
\end{minipage} 
\end{figure}

The detector is based on a new type of Resistive Plate Chamber (RPC) with Diamond-Like Carbon (DLC) electrodes sputtered on thin Kapton films as shown in Fig.~\ref{fig:DetectorConcept}.
The overall material thickness can be kept below $0.1\%\,X_0$ with four-layer configuration.
A single-layer efficiency as high as 40\,\% is required for the overall efficiency of 90\,\%.

\section{Single layer prototype}
A single-layer prototype with $2\,\mathrm{cm}\times 2\,\mathrm{cm}$ size is constructed as shown in Fig.~\ref{fig:PrototypePicture}.
The gap of $384\,\mathrm{\mu m}$-thick is made by pillars formed with a photolithography technique.
The gap is filled with a gas mixture of 94\,\% Freon (R134a), 5\,\% iso-$\mathrm{C_4H_{10}}$ and 1\,\% $\mathrm{SF_6}$.
The high voltage is supplied by conductive tape attached on the edge of the DLC surface.
The readout is implemented with a 1\,cm wide aluminum strip. 
The signal is readout from both ends of the strip, then amplified by a 38\,dB amplifier and fed into the DRS4 waveform digitizer \cite{RITT2004470}.

\section{Performance measurement}
Three different measurements were performed to evaluate the performance for the MIP positron and low-momentum muon.
Firstly, the MIP detection efficiency and timing resolution were measured at low rate, as described in Sec.~\ref{sec:MIPLowRate}.
Secondly, the RPC's response to a low-momentum muon was measured at low rate, as described in Sec.~\ref{sec:MuLowRate} with discussion on the results.
Finally, the performance in a high-rate low-momentum muon beam was measured, as described in Sec.~\ref{sec:MuHighRate}.

\subsection{Efficiency and timing resolution for MIP positrons at low rate}\label{sec:MIPLowRate}
\begin{figure}[tbp]
\centering
\includegraphics[width=0.4\linewidth]{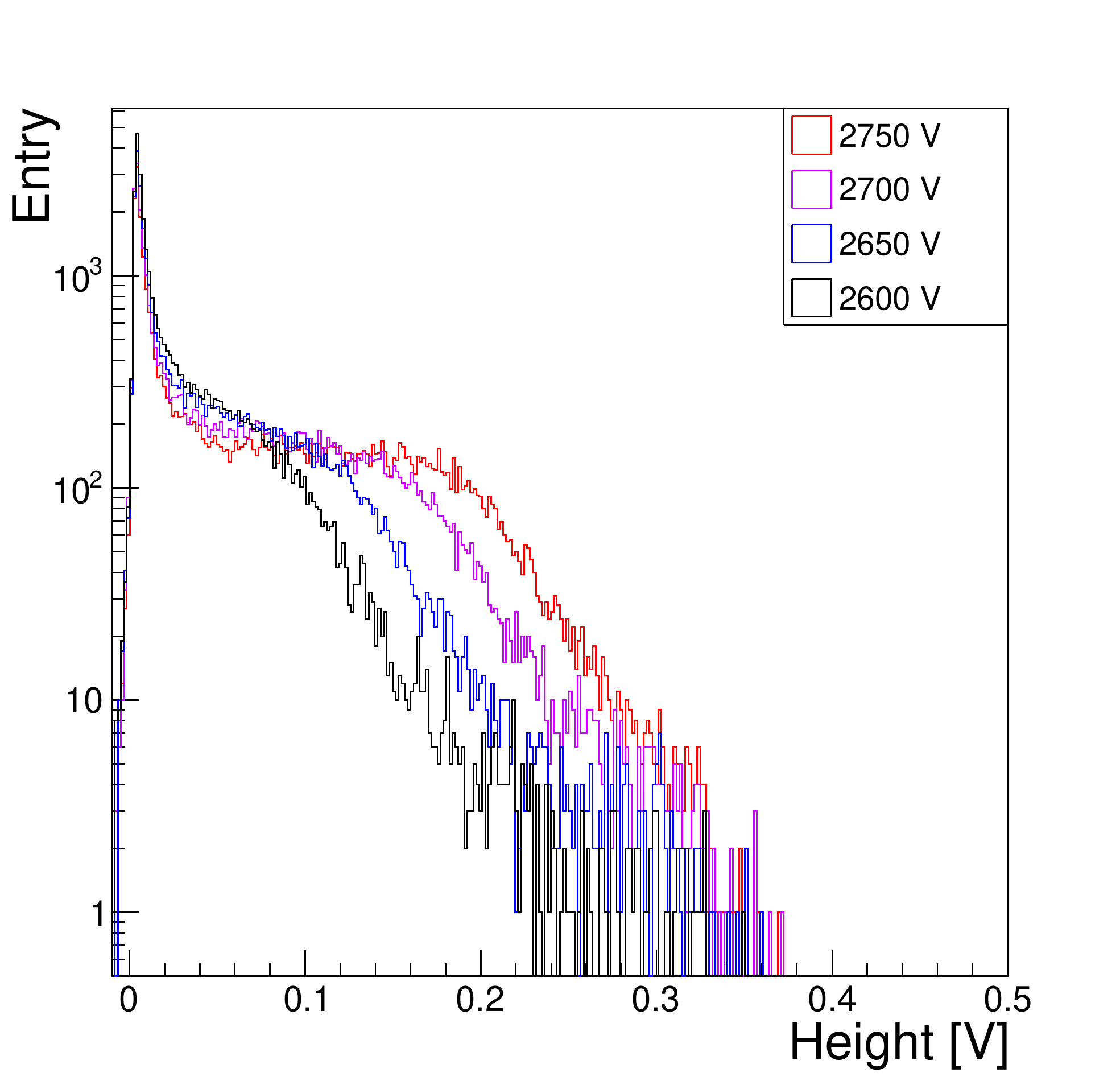}
\caption{\label{fig:LowRateMichelSpectra}Signal height distributions obtained for MIP positrons at low rate.}
\end{figure}

The efficiency and the timing resolution were measured with positrons from muon decay ($\sim10\,\mathrm{kHz/cm^2}$).
The signal height spectra are shown in Fig.~\ref{fig:LowRateMichelSpectra} changing the applied voltage between 2.6--2.75\,kV.
The detection efficiency was measured to be 60\,\% with a 10\,mV signal threshold at 2.75\,kV, which satisfies the requirement ($>40\,\%$).
The timing resolution was evaluated to be 170\,ps at 2.75\,kV, which also well satisfies the requirement ($<1\,\mathrm{ns}$).

\subsection{Response to low-momentum muon at low rate}\label{sec:MuLowRate}
\begin{figure}[tbp]
\centering
\begin{minipage}{0.45\linewidth}
\includegraphics[width=0.7\linewidth]{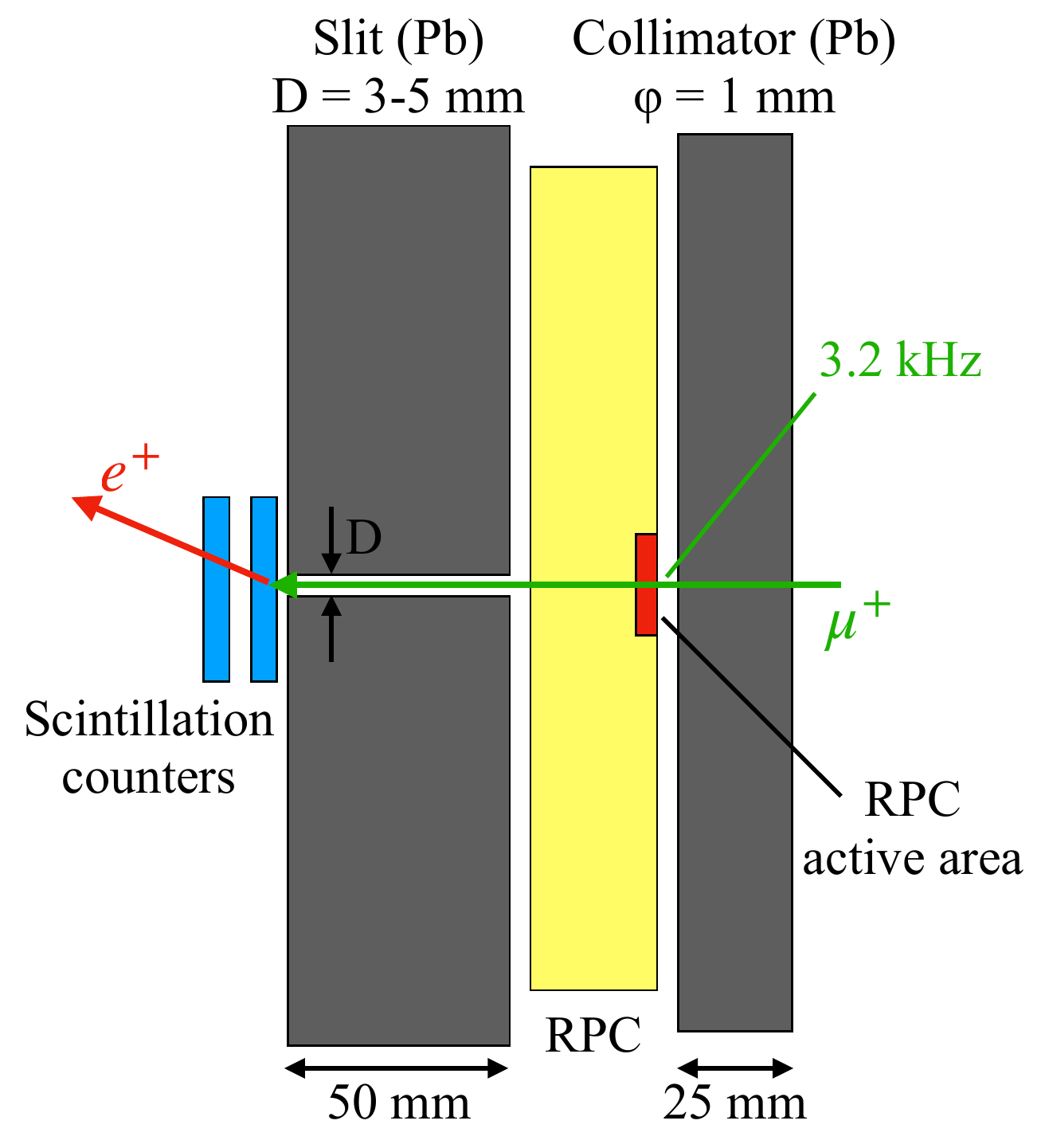}
\caption{\label{fig:setup_muon1}Schematic view of the setup for the measurement with low-momentum muon at low rate}
\end{minipage}\hspace{0.05\linewidth}
\begin{minipage}{0.4\linewidth}
\centering
\includegraphics[width=\linewidth]{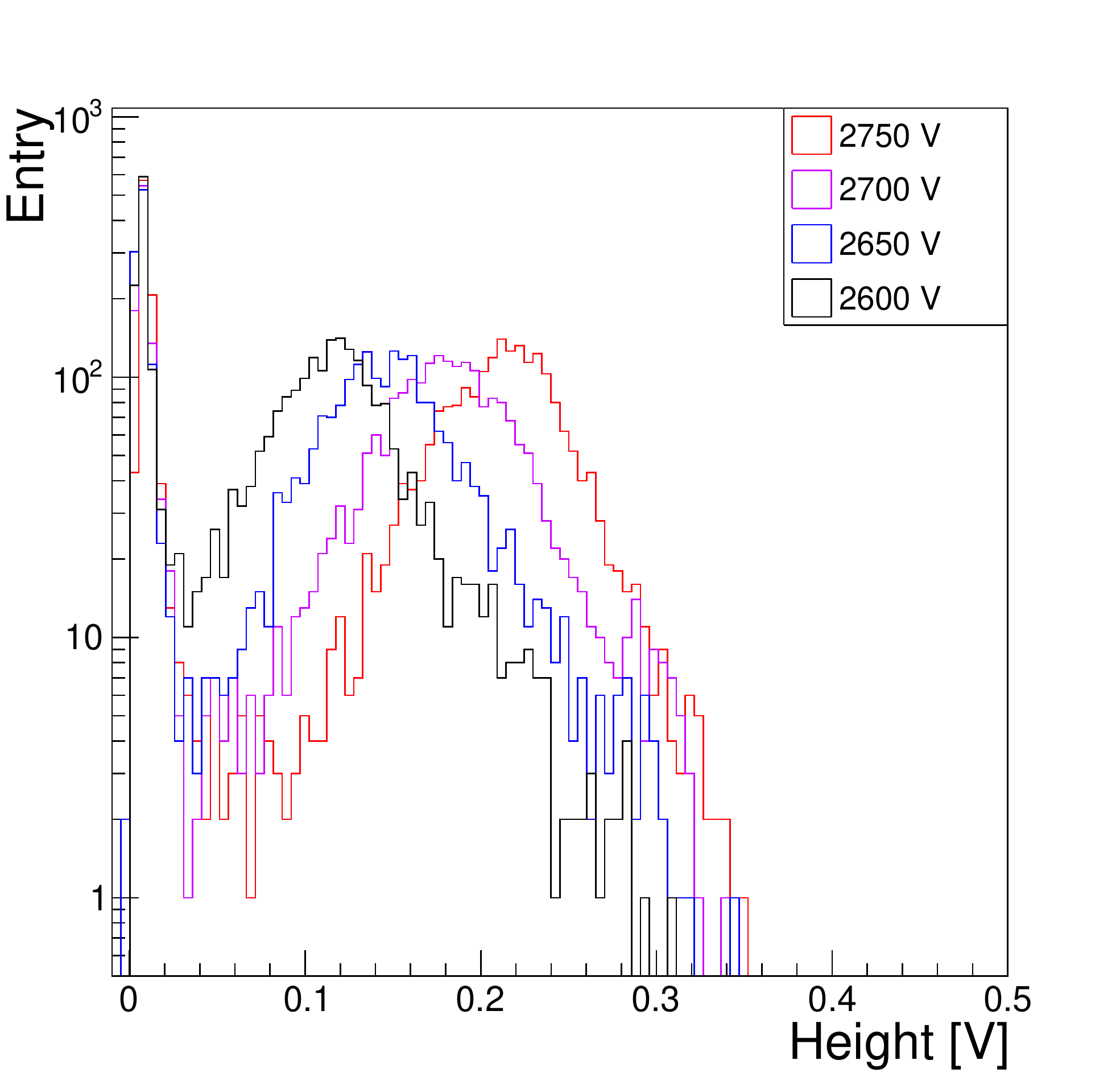}
\caption{\label{fig:LowRateMuonSpectra}Signal height distributions obtained for low-momentum muon at low rate.}
\end{minipage}
\end{figure}

The response to a low-momentum muon (28\,MeV/$c$) was measured with the muon beam at the $\mathrm{\pi E5}$ beam line of Paul Scherrer Institute.
A collimated muon beam was injected at 3.2\,kHz into the prototype as shown in Fig.~\ref{fig:setup_muon1}.
To avoid triggering positron hits, events with both muon and subsequent (within 100--450\,ns from muon hit) decay positron emitted toward downstream were selected with two scintillation counters.
The pulse height spectra for different voltages between 2.6--2.75\,kV are shown in Fig.~\ref{fig:LowRateMuonSpectra}, where the main peak around 0.1--0.2\,V corresponds to the muon events.
The pedestal peak is likely due to the accidental triggers by background radiations. 
This background prevents us from measuring the efficiency for low-momentum muon although it is found to be higher than 75\,\% at 2.75\,kV from the event fraction above the pedestal.

Even though the energy deposit by a low-momentum muon is an order of magnitude larger than that by a MIP positron, Fig.~\ref{fig:LowRateMuonSpectra} shows that the signal height is around the same level as that for the MIP positrons.
This can be explained by the space charge effect, which saturates the evolution of avalanche charge when the total avalanche gain reaches $\mathrm{O(10^7 \textrm{--} 10^8)}$ \cite{LIPPMANN200454}.

The rate capability is determined by the gain reduction due to the voltage drop caused by the signal current flowing on the resistive electrodes \cite{Aielli_2016}.
The observed saturation for the low-momentum muon is, therefore, even advantageous when the RPC is operated in a low-momentum muon beam at high rate as in MEG~II since the voltage drop can be suppressed.

\subsection{Performance in low-momentum muon beam at high rate}\label{sec:MuHighRate}
\begin{figure}[tbp]
\centering
  \begin{minipage}{0.28\linewidth}
  \includegraphics[width=\linewidth]{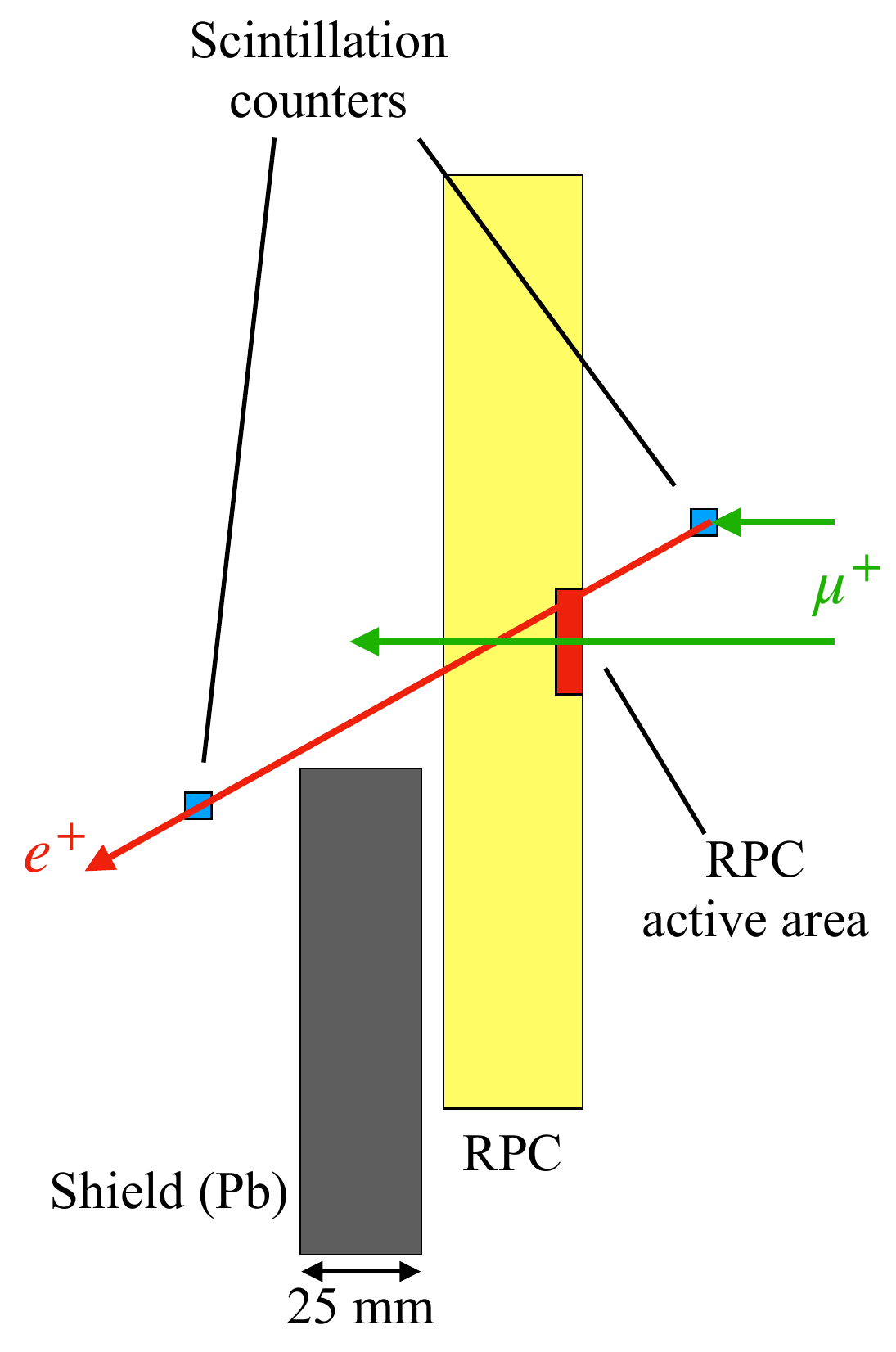}
  \end{minipage}\hspace{0.05\linewidth}%
  \begin{minipage}{0.5\linewidth}
  \includegraphics[width =\linewidth]{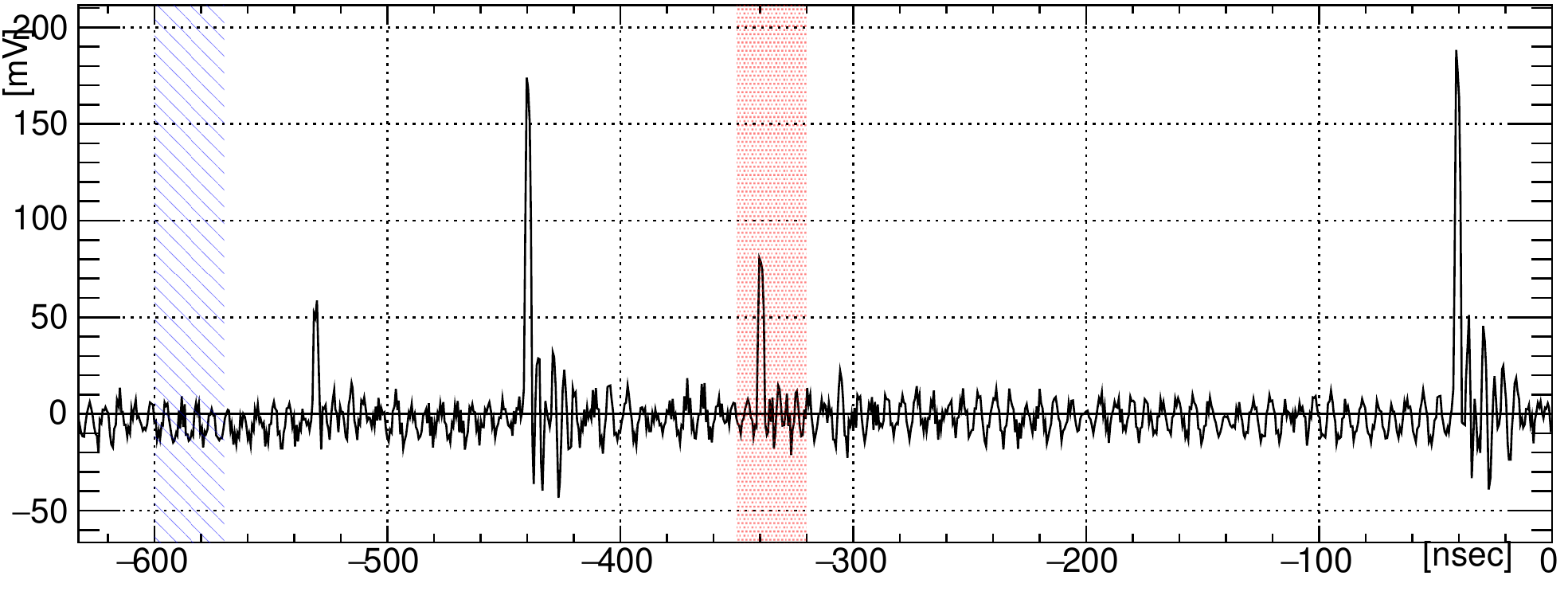}
  \end{minipage}
  \caption{\label{fig:HighRateSetup}Schematic view of the setup of measurement with high rate muon beam and a typical event display. The on-timing (off-timing) window is depicted as the red hatched (blue shaded) region.}
\end{figure}

The detection efficiency for MIP positrons in a high-intensity low-momentum muon beam was measured at 2.75\,kV with a setup shown in Fig.~\ref{fig:HighRateSetup}.
In this measurement, the RPC was exposed to a muon beam with a spread of $(\sigma_{x},\sigma_{y}) = (13\,\mathrm{mm},23\,\mathrm{mm})$ at a rate of $1\,\mathrm{MHz/cm^2}$ at the center.
The readout region of the RPC was aligned to the beam center.

Two scintillation counters were aligned to select positrons from decay of muon stopped in the upstream counter. 
The angle between the flight path of the triggered positron and the beam axis was set to $30^\circ$, which is close to the average incident angle of the positrons from the radiative muon decay to the RPC in the MEG~II experiment.

\begin{figure}[tbp]
\centering
\includegraphics[width=0.4\linewidth]{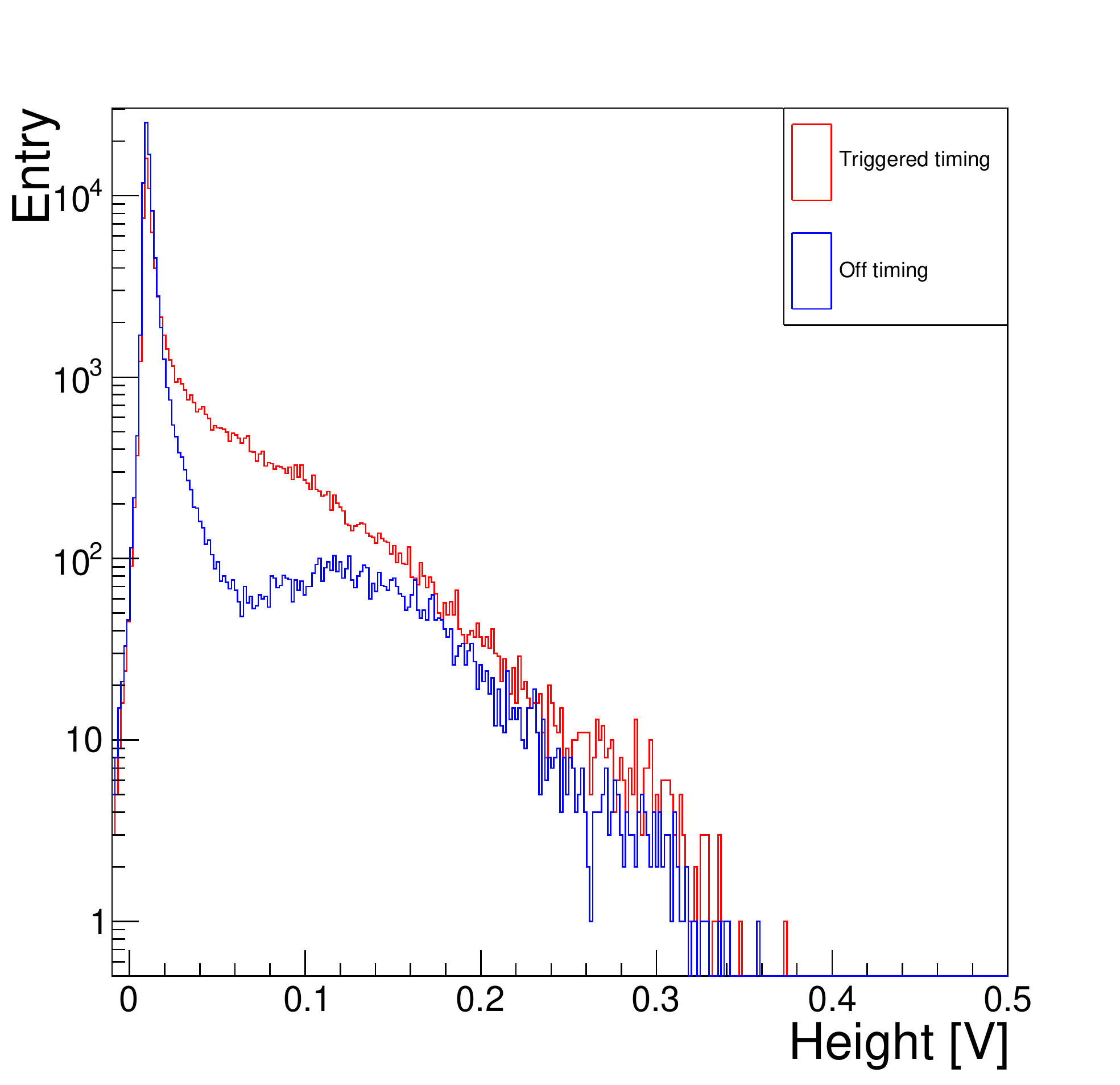}
\caption{\label{fig:HighRateSpectra}The signal height distribution obtained at 2.75\,kV in test with a high rate low-momentum muon beam. The red (blue) histogram shows the distribution in the on-timing (off-timing) window.}
\end{figure}

A typical event display is also shown in Fig.~\ref{fig:HighRateSetup}.
The waveform data were analyzed both for the on-triggered timing and off-timing windows with 30\,ns width each.
In the on-timing window, the positron events are expected and the signal height spectrum is shown as the red histogram in Fig.~\ref{fig:HighRateSpectra}, where the efficiency is evaluated to be 35\,\%.
The signal height spectrum in the off-timing window is shown as blue histogram in Fig.~\ref{fig:HighRateSpectra}, where the bump around 0.1\,V comes from the accidental muon events.

The voltage drop due to the high signal current at high rate was estimated by comparing the signal height spectra in this measurement (at 2.75\, kV) with those measured in the low rate measurements (at different voltages) in Sec.~\ref{sec:MIPLowRate} and Sec.~\ref{sec:MuLowRate}.
Both the muon and MIP spectra appear similar to those obtained at low rate at 2.6--2.65\,kV, which suggests a voltage drop of 100--150\,V.
Though $\sim 150\,\mathrm{V}$ drop is too large to achieve 40\,\% single layer efficiency, it can be reduced with further design optimization as will be discussed in Sec.~\ref{sec:ExpectedPerformance}.

The expected voltage drop can also be calculated from the geometry and the resistivity ($\sim 60\,\mathrm{M\Omega/sq}$ for the prototype) of the DLC, the measured hit rate and the avalanche charge.
The calculation also gives a voltage drop of 100--150\,V, which agrees with the estimation from the observed signal height distribution.
This result confirms that the performance in a high-rate beam can be projected from the voltage drop calculation.

\section{Expected performance in MEG~II experiment}\label{sec:ExpectedPerformance}
\begin{figure}[tbp]
\centering
\includegraphics[width=0.35\linewidth]{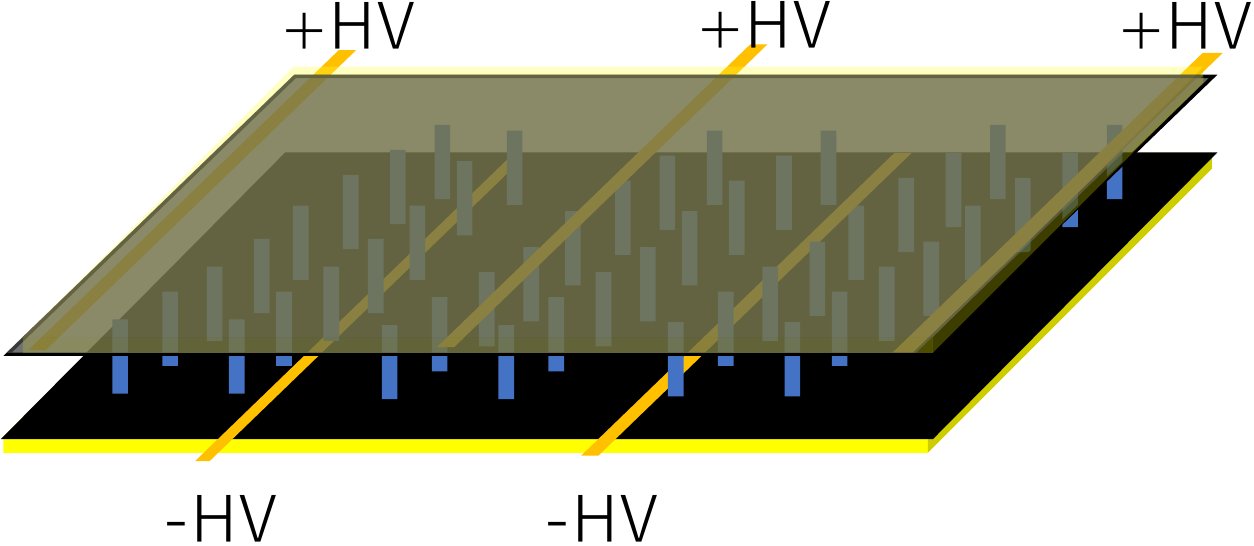}
\caption{\label{fig:StripGeometry}Schematic explaining the structure of multi-strip high voltage supply.}
\end{figure}

The full-scale RPC for the MEG II experiment should be 20\,cm in diameter.
It would suffer from a much larger voltage drop due to the larger area of the DLC.
In addition, the beam profile of $4\,\mathrm{MHz/cm^2}$ at the center and spread of $\sigma_x =\sigma_y = 20\,\mathrm{mm}$ are harsher than that in the prototype test.
The voltage drop can be mitigated with a multi-strip HV supply as shown in Fig.~\ref{fig:StripGeometry}.
The voltage drop can be reduced by the small strip pitch independently of the total detector size, although the pitch cannot be too small because of the increase of the inactive area.

With this concept, the voltage drop can be kept at a reasonable level of 100\,V for the applied voltage of 2.75\,kV with the strip pitch of  1\,cm and the surface resistivity of the DLC of $10\,\mathrm{M\Omega/sq}$.
This would allow us to operate the RPC with the required performance in the high-rate muon beam of the MEG II experiment.

We still, however, have several technical challenges to be solved to realize this design.
Firstly, the implementation of the conductive strips on the DLC surface is not trivial.
Secondly, the control of the surface resistivity of the DLC is not easy.
We currently suffer a large fluctuation by a factor 3--5.
Finally, tests with $10\,\mathrm{M\Omega/sq}$ have not been performed because of the issue of resistivity control.
A stable operation must be achieved with $10\,\mathrm{M\Omega/sq}$ resistivity while $>30\,\mathrm{M\Omega/sq}$ resistivity has been used in the measurements so far.

\section{Conclusion}
A new type of RPC with DLC electrodes is under development for the MEG~II experiment, where MIP detection in a high-rate low-momentum muon beam is required.
Using $2\,\mathrm{cm}\times 2\,\mathrm{cm}$ size prototype, the performance of RPC was measured for MIP positron and low-momentum muon beam with rate up to $1\,\mathrm{MHz/cm^2}$.
The results show that 90\,\% overall efficiency is achievable with 4-layer configuration even in the high-intensity muon beam of the MEG~II experiment, while there still remain some technical challenges.
Our results are encouraging for possible application to other experiments where the RPCs have not been available due to the limitted rate capability of the conventional glass-based RPCs.

\section{Acknowledgments}
This work is supported by JSPS Core-to-Core Program, A. Advanced Research Networks JPJSCCA20180004.

\section*{References}
\bibliography{reference}

\providecommand{\newblock}{}
\begin{thebibliography}{1}
\expandafter\ifx\csname url\endcsname\relax
  \def\url#1{{\tt #1}}\fi
\expandafter\ifx\csname urlprefix\endcsname\relax\def\urlprefix{URL }\fi
\providecommand{\eprint}[2][]{\url{#2}}

\bibitem{MEGIIdesign}
Baldini A~M {\em et~al.\/} 2018 {\em Eur. Phys. J. C\/} {\bf 78} 380
  \urlprefix\url{https://doi.org/10.1140/epjc/s10052-018-5845-6}

\bibitem{RITT2004470}
Ritt S 2004 {\em Nuclear Instruments and Methods in Physics Research Section
  A\/} {\bf 518} 470--471
  \urlprefix\url{https://www.sciencedirect.com/science/article/pii/S016890020302922X}

\bibitem{LIPPMANN200454}
Lippmann C and Riegler W 2004 {\em Nuclear Instruments and Methods in Physics
  Research Section A: Accelerators, Spectrometers, Detectors and Associated
  Equipment\/} {\bf 517} 54--76 ISSN 0168-9002
  \urlprefix\url{https://www.sciencedirect.com/science/article/pii/S0168900203026421}

\bibitem{Aielli_2016}
Aielli G {\em et~al.\/} 2016 {\em Journal of Instrumentation\/} {\bf 11}
  P07014--P07014 \urlprefix\url{https://doi.org/10.1088/1748-0221/11/07/p07014}

\end{thebibliography}

\end{document}